\documentclass[aps,prl,preprint,superscriptaddress,showpacs,showkeys]{revtex4-1}

\usepackage{amsmath}
\usepackage{amsfonts}
\usepackage{amssymb}
\usepackage{bm}

\begin{document}


\title{Spin states of Dirac equation \\and Rashba spin-orbit interaction}


\author{A. A. Eremko}
\email[]{eremko@bitp.kiev.ua}
\affiliation{Bogolyubov Institute for Theoretical Physics, \\
Metrolohichna str., 14b, Kyiv, 03680, Ukraine}
\noaffiliation
\author{L. S. Brizhik}
\email[]{brizhik@bitp.kiev.ua}
\altaffiliation{}
\affiliation{Bogolyubov Institute for Theoretical Physics, \\
Kyiv,  Ukraine}
\author{V. M. Loktev}
\email[]{vloktev@bitp.kiev.ua}
\affiliation{Bogolyubov Institute for Theoretical Physics, \\
 Kyiv,  Ukraine}
\affiliation{National Technical University of Ukraine "KPI", \\
 Kyiv, Ukraine
}


\date{\today}

\begin{abstract}
The problem of the spin states corresponding to the solutions of Dirac equation is studied. In particular, the three sets of the eigenfunctions of Dirac equation are obtained. In each set the wavefunction is at the same time the eigenfunction of one of the three spin operators, which do not commute with each other, but do commute with the Dirac Hamiltonian. This means that the eigenfunctions of Dirac equation describe three independent spin states. The energy spectrum is calculated for each of these sets for the case of quasi-two-dimensional electrons in a quantum well. It is shown that the standard Rashba spin-orbit interaction takes place in one of such states only. In another one this interaction is not formed at all, and for the third one it leads to the band spectrum which is anisotropic in the plane domain of the propagation of a free electron and is  different from the isotropic spectrum of Rashba type.  

\end{abstract}


\maketitle


\section{\label{I}Introduction}
Exploiting the spin degree of freedom of charge carriers (electrons or holes) seemed impossible up to relatively
recent times. At present it is the main aim of the new branch of electronics,
called "spintronics" \cite{Fabian}. Controlling particle spin is possible because of the spin-orbit 
interaction (SOI), which binds the spin and momentum of a particle in an external inhomogenous 
potential. Theoretical and experimental study of this problem is very important, in particular, for 
the study of layered semiconducting heterostructures, or quantum wells
(QW), in which charge carriers captured by the potential of some narrow layer, represent
practically two-dimensional (2D) electron gas. In the presence of the anisotropy determined by the applied electric field and/or special configuration of
the heterostructure, 
when the anisotropy is transversal to this layer, 
SOI removes spin the degeneracy of the 2D conducting or valent band. This
phenomenon is known as Rashba effect \cite{Rashba1, Rashba2}.

A plane potential well is one of the mostly used models to study the properties of quasi-2D
electron gas. Nonrelativistic problem is solved for the electron propagation in the external
potential which is inhomogenous in one direction and homogenous in two other directions, taking into account the potential relief. If the potential has the form of the well
(in the simplest case of the rectangular potential box) in the direction of inhomogenuity,
electrons are captured by this potential layer. They preserve the momentum in the
plane of the layer and fill in discrete energy levels of the spacial quantization,
whose number is determined by the depth of the potential well. In the result, such layer
(interface) is characterized by the set of the 2D electron bands. In the nonrelativistic approximation the dispersion laws of these bands are determined by the solutions of the Schr\"{o}dinger
equation which contains SOI. As it was mentioned above, SOI determines spin orientation. In the case of an  asymmetric QW, it determines Rashba dispersion law, or band splitting
in the $\bf{k}$-space depending on spin direction.

It is well known that the existence of the electron spin is a natural consequence
of the relativistic Dirac equation (DE) \cite{Dirac}, and that SOI in Shr\"{o}dinger equation 
results from the expansion of DE with respect to the degrees of $ 1/c $ \cite{Bete, LandauIV, Davydov}. A particle wavefunction $ \Psi (\mathbf{r}) $  in  Dirac quantum
theory is a four-component coordinate function which is represented in the form of a matrix which contains one-column (bispinor) coordinate functions. In this theory the operators related
to the spin of a particle, are sixteen $4 \times 4$ Dirac matrices, including the unit matrix. In the
nonrelativistic approximation bispinor wavefunction is approximated by a spinor one. In this
case the operators of the spin projections are given by the three Pauli matrices. Usually it is assumed that
Shr\"{o}dinger equation which includes SOI, takes into account all possibilities of DE. Nevertheless, the question, how completely and exactly nonrelativistic approximation describes
the spin degree of freedom, remains open. This problem is of special importance at present when the 
perspectives of exploring particle spin in the processes of information storage and transfer, become realistic.

In the present paper we solve DE with asymmetric rectangular potential which models a QW. 
Dirac equation with rectangular (step-like or box-like) potential was studied in various papers starting with the famous paper of Klein \cite{Klein}. Mostly, these papers studies 1D electron propagation, perpendicular to the rectangular step, with the main attention to the role of the relativistic effects \cite{Klein, Dombey, Krekora} or to the role of specific potentials which capture particles.
In particular, in  \cite{Gil} this potential was represented in the form of a spatial dependence of a mass of a relativistic particle, and in  \cite{Leo} a quaternion potential was studied. 

Below we study the spin states of quasi-2D, due to the presence of a QW, electrons and calculate their energies as functions of their 2D wave vectors.  The corresponding solutions are analyzed for the values of energies relevant to important problems of solid state physics, and compared with the results of nonrelativistic model. For the readers convenience and completeness of the paper we give some known results from the theory of DE. 

\section{\label{DirEq}Dirac equation}

Stationary DE for a particle in the absence of the magnetic field has the form \cite{Dirac, Bete, LandauIV}
\begin{equation}
\label{DE}
\left[ c\bm{\hat{\alpha}}  \mathbf{p}  + V(\mathbf{r}) + \hat{\beta} m c^{2} \right] \Psi = E \Psi \,.
\end{equation}
Here $ c $ is the speed of the light, $ m $  is the mass of a particle, $ \mathbf{p} = -i\hbar \bm{\nabla} $ is its momentum operator, $ \bm{\hat{\alpha}} = \sum_{j} \mathbf{e}_{j} \hat{\alpha}_{j} $,   $ \hat{\alpha}_{j} $ ($ j=x,y,z $) and $ \hat{\beta} $ are Dirac matrices,  $V(\mathbf{r})$ is the external potential, and  $ \Psi (\mathbf{r}) $ is a four-component coordinate function introduced above. Dirac matrices satisfy the following conditions   \cite{Bete, LandauIV, Davydov}
\begin{eqnarray}
\label{DMrel}
\hat{\alpha}_{j}^{2} =\hat{\beta}^{2} =  \hat{I} , \nonumber \\
\hat{\alpha}_{j} \hat{\beta} + \hat{\beta} \hat{\alpha}_{j} = 0 ,\\
\hat{\alpha}_{j} \hat{\alpha}_{k} + \hat{\alpha}_{k} \hat{\alpha}_{j} = 0 \:(j \neq k) , \nonumber
\end{eqnarray} 
where $ \hat{I} $ is a  unit $ 4 \times 4 $ matrix. Using Dirac matrices, it is possible to form sixteen independent matrices which form the group in the sense that the product of any two matrices is a matrix which belongs to this set, with the accuracy of the constant coefficient equal to $ \pm 1 $ or $ \pm i $ (see \cite{LandauIV, Bete}). Below we write down these Hermitian conjugate Dirac matrices in the standard representation, which we will use below:  
\[  
\hat{I} = \left( \begin{array}{cc}
\hat{I}_2 & 0 \\ 0 & \hat{I}_2
\end{array}  \right) \, ,\:\hat{\rho }_{1} = \left( \begin{array}{cc}
0 & \hat{I}_2 \\ \hat{I}_2 & 0
\end{array}  \right) \,,\:\hat{\rho}_{2} = \left( \begin{array}{cc}
0 & -i \hat{I}_2 \\ i \hat{I}_2 & 0
\end{array}  \right) \,,\:\hat{\rho}_{3} = \left( \begin{array}{cc}
\hat{I}_2 & 0 \\ 0 & -\hat{I}_2
\end{array}  \right) \,,
\]
\begin{eqnarray}
\label{Dm}
\bm{\hat{\alpha}} = \left( 
\begin{array}{cc}
0 & \bm{\hat{\sigma}} \\ \bm{\hat{\sigma}} & 0
\end{array} \right) \, ,\:\bm{\hat{\Sigma}} = \left( 
\begin{array}{cc}
\bm{\hat{\sigma}} & 0 \\ 0 & \bm{\hat{\sigma}}
\end{array} \right) \, ,\\ 
\bm{\hat{\Gamma}} = \left( 
\begin{array}{cc}
0 & -i\bm{\hat{\sigma}} \\ i\bm{\hat{\sigma}} & 0
\end{array} \right) \, ,\:\bm{\hat{\Omega}} = \left( 
\begin{array}{cc}
\bm{\hat{\sigma}} & 0 \\ 0 & -\bm{\hat{\sigma}}
\end{array} \right) .\nonumber
\end{eqnarray}
Here $ \hat{I}_2 $ is a unit $ 2\times 2 $ matrix, $ \hat{\sigma}_{j} $ ($ j=x,y,z $) are Pauli matrices which satisfy the following relations: 
\[ 
\hat{\sigma}_{x} \hat{\sigma}_{y} = i\hat{\sigma}_{z} ,\quad  \hat{\sigma}_{z} \hat{\sigma}_{x} = i\hat{\sigma}_{y} , \quad \hat{\sigma}_{y} \hat{\sigma}_{z} = i\hat{\sigma}_{x},\quad  \hat{\sigma}_{j}^{2} = \hat{I}_2 .
\]
Pauli matrices are usually written in the form
\begin{equation}
\label{PM}
\hat{\sigma}_{x} = \left( 
\begin{array}{cc}
0 & 1 \\ 1 & 0
\end{array} \right) , \quad \hat{\sigma}_{y} = \left( 
\begin{array}{cc}
0 & -i \\ i & 0
\end{array} \right) , \quad \hat{\sigma}_{z} = \left( 
\begin{array}{cc}
1 & 0 \\ 0 & -1
\end{array} \right) ,
\end{equation}
where $ z $-axis defines a chosen direction. In (\ref{Dm}) the notations of $ \hat{\rho}_{j} $ matrices  are used as in  \cite{Sokolov} with $ \hat{\rho}_{3} \equiv \hat{\beta} $, as it is used in the relativistic electron theory. 

The very notion of the spin degree of freedom is related to the eigen momentum of a particle, called 'spin'.
Spin is the consequence of the fact that, according to DE, in an isotropic space  an integral of motion is the total momentum  
\begin{equation}
\label{J_tot}
\mathbf{\hat{J}} = \mathbf{L} + \mathbf{S} \, .
\end{equation}
Here $ \mathbf{L} = \mathbf{r} \times \mathbf{p} $ is the operator of the orbital momentum, and $ \hat{\mathbf{S}} = (\hbar /2) \bm{\hat{\Sigma}} $ is the spin operator with the projections of the vector $\bm{\hat{\Sigma}} $ defined in (\ref{Dm}). The square of the modulus  of the spin $ \hat{\mathbf{S}}^{2} $ is an integral of motion and corresponds to the eigenvalue   $ s = 1/2 $. Therefore, the spin projection $ \hbar m_s $ on some chosen direction takes two values which correspond to the magnetic quantum number  $ m_s = \pm 1/2 $. Nevertheless, it is impossible to relate eigen spinors to this quantum number since the operator  $ \hat{\mathbf{S}}$ does not commute with the Hamiltonian, and, hence, does not have a system of eigenfunctions which is joint with it.  Therefore, to define the spin eigenstates of relativistic particles, one has to find such spin operators which commute with the corresponding Hamiltonian. 

The choice  of the spin operator which determines spin eigenstates, is not unique in the case of a free particle, since there  are several operators that commute with the Hamiltonian. One of them is the operator of the spin momentum projection  parallel to the momentum or to the direction of motion:
\begin{equation}
\label{S_p}
\hat{S}_{\parallel} = \bm{\hat{\Sigma}}  \mathbf{p} \, .
\end{equation}
Operators of vector momentum,   $ \mathbf{p} $, and of total momentum,    $ \mathbf{J} $, are integrals of free motion, as well as some other vector operators, which commute with the Hamiltonian of Eq. (\ref{DE}):
\begin{equation}
\label{eps}
\bm{\hat{\epsilon}} = \bm{\hat{\Omega}} \times \mathbf{p} \, ,
\end{equation}
\begin{equation}
\label{mu}
\bm{\hat{\mu}} = \bm{\hat{\Sigma}} + \frac{\bm{\hat{\Gamma}} \times \mathbf{p}}{mc} \, ,
\end{equation}
\begin{equation}
\label{genS}
\bm{\hat{\mathcal{S}}} = \bm{\hat{\Omega}} + \hat{\rho}_{1} \frac{\mathbf{p}}{mc} \, .
\end{equation}
In \cite{Sokolov} the components of the vector (\ref{eps}) are treated as the components of electric spin polarization, of the vector (\ref{mu}) -- as components of magnetic spin polarization, and the 3D vector  (\ref{genS}) together  with the value (\ref{S_p}) – as the components of 4-pseudovector of spin polarization. For a free motion of a particle one can also introduce the operator of spin polarization 
\[\bm{\hat{s}}_{0} = \frac{\mathbf{p} \hat{S}_{\parallel} + \left[ \mathbf{p} \left[ \bm{\hat{\mathcal{S}}} \mathbf{p} \right] \right] }{p^{2}} \, . \]
The latter allows to describe spin projections on the arbitrary direction  $ \mathbf{e} \cdot \mathbf{s}_{0} $ where $ \mathbf{e} $ is an arbitrary unit vector. 

\subsection{\label{QWP}Quantum well potential} 

Below we consider the simplest case of a 1D potential, or QW. Let us choose the direction of   $ z$-axis along the normal to the QW plane. The potential created by such a well, depends on one coordinate only,  $ V(\mathbf{r}) = V(z) $. In this case the state of a particle with a given value of a 2D momentum in the $ xy$-plane,  $\mathbf{k}=(k_x, k_y) $, is described by the function  
\begin{equation}
\label{Psi_2D}
\Psi_{\mathbf{k}}(\mathbf{r}) = e^{i(k_{x} x + k_{y} y)} \varPsi (z) \, .
\end{equation}
Substituting (\ref{Psi_2D}) in DE (\ref{DE}), we can rewrite it in the form 
\[  
\left[ -i\hbar c\hat{\alpha}_{z} \frac{d}{dz} + \hbar c \mathbf{k} \bm{\hat{\alpha}} + V(z)\hat{I} + m c^{2} \hat{\beta} \right] \varPsi (z) = E \varPsi (z) \, .
\]

Representing the bispinor  $ \varPsi (z) $ in the form 
\begin{equation}
\label{bispin}
\varPsi (z) = { \psi (z) \choose \varphi (z) } ,
\end{equation}
we transform Eq. (\ref{DE}) to the following system of equations: 
\begin{eqnarray}
\label{Eqs_2D}
-i\hbar c\hat{\sigma}_{z} \frac{d}{dz} \varphi + \hbar c \mathbf{k} \bm{\hat{\sigma}} \varphi + \left[ mc^{2} + V(z) \right] \psi = E \psi \, , \nonumber \\
-i\hbar c\hat{\sigma}_{z} \frac{d}{dz} \psi + \hbar c \mathbf{k}  \bm{\hat{\sigma}} \psi - \left[ mc^{2} - V(z) \right] \varphi = E \varphi \, .
\end{eqnarray}

It is easy to see that in the case of an inhomogenous potential $ V(z)\neq \mathtt{const} $ not all spin operators  (\ref{S_p})-(\ref{genS}) commute with the Hamiltonian. Only the following components of the vector operators   (\ref{mu})-(\ref{genS}) are conserved:
\begin{equation}
\label{eps_z}
\hat{\epsilon}_{z} = \hat{\Omega}_{x} p_{y} - \hat{\Omega}_{y} p_{x} \, ,
\end{equation}
\begin{equation}
\label{mu_z}
\hat{\mu}_{z} = \hat{\Sigma}_{z} + \frac{1}{mc} \left( \hat{\Gamma}_{x} p_{y} - \hat{\Gamma}_{y} p_{x} \right) \, ,
\end{equation}
\begin{equation}
\label{S_x_y} 
\hat{\mathcal{S}}_{x} = \hat{\Omega}_{x} + \hat{\rho}_{1} \frac{p_{x}}{mc} , \quad \hat{\mathcal{S}}_{y} = \hat{\Omega}_{y} + \hat{\rho}_{1} \frac{p_{y}}{mc} \, .
\end{equation}
Since these operators commute with the Hamiltonian but do not commute with each other, the system of the  eigenfunctions of the Hamiltonian can be joint with one of them, only. Therefore, the solution of Eqs. (\ref{Eqs_2D}) can be represented in the form of the three sets of eigen wavefunctions, with different physical meaning of the spin quantum numbers.  Worth mentioning also that since the projections  $\hat{\mathcal{S}}_{x}  $ and  $ \hat{\mathcal{S}}_{y} $ do not commute, the anisotropy in the originally isotropic $ xy $-plane automatically appears with an arbitrary axis of the anisotropy. 

\subsubsection{\label{ESP} Eigenfunctions of the operator  of the electric spin polarization projection
}
\noindent First, let us consider the case when the eigenfunctions of DE are also the eigenfunctions of the operator  (\ref{eps}). This means that we search the solution of the system of equations  (\ref{Eqs_2D}), which is joint with the equation 
\[ \hat{\epsilon}_{z} \Psi_{\mathbf{k}}(\mathbf{r}) = \epsilon \Psi_{\mathbf{k}}(\mathbf{r}) \, . \]
Substituting in it the function (\ref{Psi_2D}) and taking into account Eq. (\ref{bispin}), the explicit form of the operator  $ \hat{\epsilon}_{z} $ and the corresponding Dirac matrices (\ref{Dm}), we get the equation 
\begin{equation}
\label{Eq_epsil}
-\hbar k \left( 
\begin{array}{cc}
\hat{\Lambda}_R & 0 \\ 0 & -\hat{\Lambda}_R
\end{array} \right) { \psi (z) \choose \varphi (z) } = \epsilon { \psi (z) \choose \varphi (z) } \, ,
\end{equation}
where according to the definition,  $ k = \sqrt{k_{x}^{2} + k_{y}^{2}} $ and 
\begin{equation}
\label{Lambda}
\hat{\Lambda}_R =
\left( \begin{array}{cc}
0 & -ie^{-i\phi} \\ ie^{i\phi} & 0
\end{array}  \right) = 
\left( \begin{array}{cc}
0 & e^{-i(\phi + \pi/2)} \\ e^{i(\phi + \pi/2)} & 0
\end{array}  \right) ,\quad   \tan \phi = \frac{k_{y}}{k_{x}} .
\end{equation}
Here the projections of the 2D wavevector are  $ k_{x} = k \cos \phi $, $ k_{y} = k \sin \phi $, and the matrix  $ \hat{\Lambda}_R $, defined as $ \hat{\Lambda}_R=(k_{x} \hat{\sigma}_{y} - k_{y} \hat{\sigma}_{x})/k $, determines Rashba SOI in the nonrelativistic approach (see \cite{Fabian,Rashba1}). 

Let us represent the eigenvalue as  $ \epsilon =-\hbar k \sigma $. They can be obtained from Eq.  (\ref{Eq_epsil}) which takes the form:
\[ \hat{\Lambda}_{R} \psi (z) = \sigma \psi (z) \, , \quad \hat{\Lambda}_{R} \varphi (z) = - \sigma \varphi (z) \, . \] 
This means that functions  $ \psi (z) $ and $ \varphi (z) $ are proportional to the eigen spinors of the matrix $ \hat{\Lambda}_R $, which are orthonormalized spinor
\begin{equation}
\label{spinors}
\chi_{+} = \frac{1}{\sqrt{2}} { e^{-i(\phi /2 + \pi /4)} \choose e^{i(\phi /2 + \pi /4)} } \, , \; 
\chi_{-} = \frac{1}{\sqrt{2}} { e^{-i(\phi /2 + \pi /4)} \choose - e^{i(\phi /2 + \pi /4)} } \, , \; 
\chi_{\sigma}^{\dagger}\chi_{\sigma '} = \delta_{\sigma ,\sigma '} \, ,
\end{equation}
that correspond to the eigenvalues  $ \sigma =  \pm 1 $, respectively, so that  $ \hat{\Lambda}_R \chi_{\pm} = \pm  \chi_{\pm} $. Therefore, the bispinor $ \varPsi (z) $ is characterized by the quantum number  $ \sigma = \pm 1 $: 
\begin{equation}
\label{varPsi_eps}
\varPsi_{\sigma} (z) = { f(z) \chi_{\sigma} \choose g(z) \chi_{-\sigma} } ,
\end{equation}
where $ f(z) $ and $ g(z) $ are arbitrary functions. We can assign the quantum number $ \sigma $ to the bispinor  (\ref{varPsi_eps}) as the quantum number that corresponds to the upper component of (\ref{varPsi_eps}). 

To find these functions, we have to substitute this bispinor into Eqs.  (\ref{Eqs_2D}), which contain matrices   $ \hat{\sigma}_{z} $ and 
\begin{equation}
\label{sigma_p} 
\mathbf{k}  \bm{\hat{\sigma}} = k_{x} \hat{\sigma}_{x} + k_{y} \hat{\sigma}_{y} = k \hat{\Lambda}(\phi ) \, , \quad \hat{\Lambda}(\phi ) = \left( \begin{array}{cc}
0 & e^{-i \phi } \\ e^{i \phi } & 0
\end{array}  \right) .
\end{equation}
Taking into account the action of these matrices on the spinors  (\ref{spinors}), namely,   $ \hat{\sigma}_{z} \chi_{\sigma} = \chi_{-\sigma} $ and $ \hat{\Lambda} (\phi )\chi_{\sigma} = i \sigma \chi_{-\sigma} $, we see that Eqs.  (\ref{Eqs_2D}) are transformed into the system of equations for functions $ f(z) $ and $ g(z) $:
\begin{eqnarray}
\label{Eqs_fg_eps}
-i\hbar c\left( \frac{d f}{d z} - \sigma k f \right) = \left[ E - V(z) + mc^{2} \right] g   \, , \nonumber \\
-i\hbar c\left( \frac{d g}{d z} + \sigma k g \right) = \left[ E - V(z) - mc^{2} \right] f   \, .
\end{eqnarray}

Below we consider the QW of the form  
\begin{equation}
\label{V_123}
V(z) = \left\lbrace \begin{array}{ll}
V_{L} & \textrm{ïðand  $ z< a $}, \\
V_{C} & \textrm{ïðand  $ a\leq z\leq b $}, \\
V_{R} & \textrm{ïðand  $ z> b $},
\end{array} \right. 
\end{equation}
where  $ V_{L} \geq V_{R} > V_{C} $ and $ d=b - a $ is the width of the QW. In other words, we assume that the space in  $z$-direction is divided into three regions in each of which the potential takes constant value  $ V_{j} = \textrm{const} $ ($ j = L,C,R $). Equations (\ref{Eqs_fg_eps}) in each of these regions are reduced to a single equation for the function  $ f_{j}(z) $:
\begin{equation}
\label{Eq_f}
- \hbar^{2} c^{2}\frac{d^{2} f_{j}}{dz^{2}} = \left[ \left( E - V_{j} \right) ^{2} - \varepsilon_{\perp}^{2}  \right] f_{j} \, .
\end{equation}
Here the notation is introduced
\begin{equation}
\label{vareps_2}
\varepsilon_{\perp} = \sqrt{ m^{2} c^{4} + \hbar^{2} c^{2} k^2 } \, .
\end{equation}
The function $ g_{j}(z) $ is determined from the second equation in  (\ref{Eqs_fg_eps}) at the given energy
\begin{equation}
\label{g_j_eps}
g_{j}(z) = - i\frac{\hbar c}{E - V_{j} + mc^{2}} \left( f'_{j} - \sigma k f_{j} \right) \, .
\end{equation}
Here and below the prime means $z$-derivative of the corresponding function. 

\subsubsection{\label{EF-MSO}Eigenfunctions of the operator 
of magnetic spin polarization projection } 
Let us now find solutions of the system (\ref{Eqs_2D}), which are joint with the equation
\[ \hat{\mu}_{z} \Psi_{\mathbf{k}}(\mathbf{r}) = \mu \Psi_{\mathbf{k}}(\mathbf{r}) \, . \]
Substituting function  (\ref{Psi_2D}) into the latter equation and taking into account Eq. (\ref{bispin}), the explicit form of the operator $ \hat{\mu}_{z} $ (\ref{mu}) and the corresponding Dirac matrices, we obtain the equation  
\begin{equation}
\label{Eq_mu}
\left( 
\begin{array}{cc}
\hat{\sigma}_{z} & i \frac{\hbar k}{mc}\hat{\Lambda}_R \\ - i \frac{\hbar k}{mc}\hat{\Lambda}_R & \hat{\sigma}_{z}
\end{array} \right) { \psi (z) \choose \varphi (z) } = \mu { \psi (z) \choose \varphi (z) } \, ,
\end{equation}
where matrix $ \hat{\Lambda}_R $ is defined in (\ref{Lambda}). 

To solve Eq. (\ref{Eq_mu}), let us chose the spinors $ \psi (z) $ and $ \varphi (z) $ in the form
\begin{equation}
\label{spinors_mu}
\psi (z) = { e^{-i\phi /2 } \psi_{1} (z) \choose e^{i\phi /2 } \psi_{2} (z) } , \quad  \varphi (z) = { e^{-i\phi /2 } \varphi_{1} (z) \choose e^{i\phi /2 } \varphi_{2} (z) }
\end{equation}
and rewrite Eq. (\ref{Eq_mu}) as follows:
\begin{eqnarray}
\left( \mu - 1 \right) \psi_{1} (z) - \frac{\hbar k}{mc} \varphi_{2} (z) = 0 \, ,\nonumber \\
- \frac{\hbar k}{mc} \psi_{1} (z) + \left( \mu + 1 \right) \varphi_{2} (z) = 0 \, ,\nonumber \\
\left( \mu + 1 \right) \psi_{2} (z) + \frac{\hbar k}{mc} \varphi_{1} (z) = 0 \, ,\nonumber \\
\frac{\hbar k}{mc} \psi_{2} (z) + \left( \mu - 1 \right) \varphi_{1} (z) = 0 \, .\nonumber 
\end{eqnarray}
From above we find the eigenvalues $ \mu = \pm \mu_{0} $, ãäå $ \mu_{0} = \sqrt{1 + \left( \hbar k/mc \right)^{2} } $. The corresponding solutions can be represented as: 
\begin{equation}
\label{mu+}
\psi_{2} (z) = - \lambda \varphi_{1} (z) , \; \varphi_{2} (z) = \lambda \psi_{1} (z), 
\end{equation}
at $ \mu = \mu_{0}$, and as 
\begin{equation}
\label{mu-}
\psi_{1} (z) = - \lambda \varphi_{2} (z) , \; \varphi_{1} (z) = \lambda \psi_{2} (z), 
\end{equation}
at $ \mu = -\mu_{0},$
with two arbitrary functions for every value of  $\mu $. Namely, $ \psi_{1} $ and $ \varphi_{1} $ at $ \mu = \mu_{0} $, and  $ \psi_{2} $ and $ \varphi_{2} $ at $ \mu = -\mu_{0} $. The coefficientò $ \lambda $ in these relations is defined as
\begin{equation}
\label{lambd}
\lambda = \frac{\hbar k}{mc\left( \mu_{0} + 1 \right) } = \frac{\hbar c k}{\varepsilon_{\perp} + mc^{2}},
\end{equation}
where $ \varepsilon_{\perp} $ is defined in   (\ref{vareps_2}). Therefore, the spinors incoming into the bispinor with the given spin quantum number  $ \mu = \pm \mu_{0} $, take the final form 
\begin{equation}
\label{VarPsi_+}
\psi_{+} (z) = { e^{-i\phi /2 } \psi (z) \choose -e^{i\phi /2 } \lambda \varphi (z) } , \quad \varphi_{+}(z)={ e^{-i\phi /2 } \varphi (z) \choose e^{i\phi /2 } \lambda \psi (z) },
\end{equation}
\begin{equation}
\label{VarPsi_-}
\psi_{-} (z) = { -e^{-i\phi /2 } \lambda \varphi \choose e^{i\phi /2 } \psi (z) } , \quad  \varphi_{-} (z) = { e^{-i\phi /2 } \lambda \psi (z) \choose e^{i\phi /2 } \varphi } .
\end{equation}
 
Substituting the above expressions for the spinors  (\ref{VarPsi_+}) or (\ref{VarPsi_-}) into (\ref{Eqs_2D}), we derive four equations for the two unknown functions  $ \psi (z) $ and  $ \varphi (z) $. Let us introduce the spin quantum number  $ \sigma $, which takes two values $ \sigma = \pm 1 $ and which defines the eigenvalues of the magnetic spin polarization  (\ref{mu_z}): $ \mu = \sigma \mu_{0} $. Then four equations (\ref{Eqs_2D}) can be written in the form
\begin{eqnarray}
i\hbar  c\sigma \frac{d \varphi}{dz} =  \hbar c k \lambda \psi - \left( E - V - mc^{2} \right) \psi \, ,\nonumber \\
i\hbar  c\sigma \lambda \frac{d \varphi}{dz} = \hbar c k \psi - \left( E - V + mc^{2} \right) \lambda \psi \, , \nonumber \\
i\hbar  c\sigma \frac{d \psi}{dz} = - \hbar c k \lambda \varphi - \left( E - V + mc^{2} \right) \varphi \, , \nonumber \\
i\hbar  c\sigma \lambda \frac{d \psi}{dz} = - \hbar c k \varphi - \left( E - V - mc^{2} \right) \lambda \varphi \, .
\nonumber 
\end{eqnarray}
Multiplying the first equation by  $ \lambda $ and extracting from it the second equation, we get the identity  $ 0=0 $. Similarly, multiplying the third equation by  $ \lambda $ and extracting from it the fourth one, we again get the same identity. Therefore, we have two equations for the two unknown functions. Let us rewrite them by adding the second equation with the first one, multiplied by  $ \lambda $, and, respectively, the fourth with the third one multiplied also by  $ \lambda $. In the result we have
\begin{eqnarray}
\label{Eqs_fg_mu}
-i\hbar  c\sigma  \frac{d \psi}{d z} = \left[ E - V(z) + \varepsilon_{\perp} \right] \varphi   \, , \nonumber \\
-i\hbar  c\sigma  \frac{d \varphi}{d z} = \left[ E - V(z) - \varepsilon_{\perp} \right] \psi   \, .
\end{eqnarray}
where the definition  (\ref{lambd}) is used. 

In the case of a rectangular QW (\ref{V_123}), Eqs.  (\ref{Eqs_fg_mu}) in each  region  are reduced to the same equation  (\ref{Eq_f}) for function $ \psi_{j} (z) $ ($ j = L,C,R $). The function $ \varphi_{j} (z) $ corresponding to the solution, has the form  
\begin{equation}
\label{g_j_mu}
\varphi_{j}(z) = - i\frac{\hbar  c\sigma}{E - V_{j} + \varepsilon_{\perp}} \psi'_{j}  \, ,
\end{equation}
which is different from (\ref{g_j_eps}). 

\subsubsection{\label{EF-SP}Eigenfunctions of the operator  of the spin polarization projection
 $ \bm{\hat{\mathcal{S}}} $ }  
As it was mentioned above, in the given configuration of the potential  $ V(z) $, the  projections of this operator in the   $xy$-plane are conserved. Since the operators of these projections do not commute between themselves, the Hamiltonian  can have the joint system of eigenfunctions, generally speaking, with any component of the operator   $ \bm{\hat{\mathcal{S}}} $,  with arbitrary direction in  $xy$-plane. In the general case this direction can be defined by a unit vector  $ \mathbf{e}_0 = \mathbf{e}_{x} \cos \phi_{0} + \mathbf{e}_{y} \sin \phi_{0} $. This means that we have to find the eigenfunctions of the following operator: 
\begin{equation}
\label{S_e}
\hat{\mathcal{S}}_{\mathbf{e}_0} \equiv \hat{\mathcal{S}}_0 = \mathbf{e}_0  \bm{\hat{\mathcal{S}}}  = \hat{\Omega}_{x}\cos \phi_{0}  + \hat{\Omega}_{y}\sin \phi_{0}  + \hbar \hat{\rho}_{1} \frac{k_{x}\cos \phi_{0} + k_{y}\sin \phi_{0} }{mc} \, .
\end{equation}
Here matrices $ \hat{\Omega}_j $ and  $\hat{\rho}_{1} $ are defined in (\ref{Dm}). Acting by the operator 
 (\ref{S_e}) on function (\ref{Psi_2D}) and taking into account (\ref{bispin}) and explicit form of Dirac matrices, gives us the following equation
\begin{equation}
\label{Eq_S}
\left( 
\begin{array}{cc}
\hat{\Lambda}(\phi _0) & \frac{\hbar k_{0}}{mc}\hat{I}_2 \\ \frac{\hbar k_{0}}{mc}\hat{I}_2 & - \hat{\Lambda}(\phi _0)
\end{array} \right) { \psi (z) \choose \varphi (z) } = s { \psi (z) \choose \varphi (z) } \, .
\end{equation}
Here the operator $\hat{\Lambda}(\phi _0)$ is defined in (\ref{sigma_p}) at $\phi =\phi _0$, and, therefore, it has the following explicit form $
\hat{\Lambda}(\phi _0) =  \cos \phi_{0}\hat{\sigma}_{x} + \hat{\sigma}_{y}\sin \phi_{0}$.
Let us represent the vector  $ \mathbf{k} $ using parallel and orthogonal components of the vector  $ \mathbf{e}_0$: $ \mathbf{k} = k_{0} \mathbf{e}_0 + k_{\perp} \mathbf{e}_{\perp} $, where $ \mathbf{e}_{\perp} = \mathbf{e}_0 \times \mathbf{e}_{z} = \mathbf{e}_{x} \sin \phi_{0} - \mathbf{e}_{y} \cos \phi_{0} $:
\begin{equation}
\label{p_e_ort}
k_{0} = k_{x} \cos \phi_{0} + k_{y} \sin \phi_{0} \, , \quad  k_{\perp} = k_{x} \sin \phi_{0} - k_{y} \cos \phi_{0} .
\end{equation}
To solve Eq. (\ref{Eq_S}), let us write the spinors $ \psi (z) $ and $ \varphi (z) $ in the form, similar to Eq. (\ref{spinors_mu}): 
\begin{equation}
\label{spinors_S}
\psi (z) = { e^{-i\phi_{0} /2 } \psi_{1} (z) \choose e^{i\phi_{0} /2 } \psi_{2} (z) } , \quad  \varphi (z) = { e^{-i\phi_{0} /2 } \varphi_{1} (z) \choose e^{i\phi_{0} /2 } \varphi_{2} (z) }.
\end{equation}
In the result, we derive the system of equations
\begin{eqnarray}
s \psi_{1} (z) - \frac{\hbar k_{0}}{mc} \varphi_{1} (z) = \psi_{2} (z) \, ,\nonumber \\
s \psi_{2} (z) - \frac{\hbar k_{0}}{mc} \varphi_{2} (z) = \psi_{1} (z) \, ,\nonumber \\
s \varphi_{1} (z) - \frac{\hbar k_{0}}{mc} \psi_{1} (z) = -\varphi_{2}(z) \, ,\nonumber \\
s \varphi_{2} (z) - \frac{\hbar k_{0}}{mc} \psi_{2} (z) = - \varphi_{1} \, .\nonumber 
\end{eqnarray}
Similarly to above, adding and extracting the first two equations, and repeating this procedure for the last ones, we obtain the system which is more convenient for analysis: 
\begin{eqnarray}
\left( s - 1 \right) \left( \psi_{1} + \psi_{2} \right) - \frac{\hbar k_{0} }{mc} \left( \varphi_{1} + \varphi_{2} \right ) = 0 \, ,\nonumber \\
- \frac{\hbar k_{0}}{mc} \left(\psi_{1} + \psi_{2} \right) + \left( s + 1 \right) \left( \varphi_{1} + \varphi_{2} \right) = 0  \, ,\nonumber \\
\left( s + 1 \right) \left( \psi_{1} - \psi_{2} \right) - \frac{\hbar k_{0}}{mc} \left( \varphi_{1} - \varphi_{2} \right) = 0 \, ,\nonumber \\
- \frac{\hbar k_{0}}{mc} \left( \psi_{1} - \psi_{1} \right) + \left( s - 1 \right) \left( \varphi_{1} - \varphi_{2} \right) = 0  \, . \nonumber
\end{eqnarray}
From here we find the eigenvalues  $ s = \pm s_{0} $, where (comp. with  (\ref{mu+})-(\ref{mu-})) $ s_{0} = \sqrt{1 + \left( \hbar k_{0}/mc \right)^{2} } $. The solutions contain two pairs of arbitrary functions which can be written in the form  
\begin{equation}
\label{s+}
\varphi_{1} (z) + \varphi_{2} (z) = \xi \left[ \psi_{1} (z) + \psi_{2} (z)\right]  , \; \psi_{1} (z) - \psi_{2} (z) = \xi \left[ \varphi_{1} (z) - \varphi_{2} (z)\right]  
\end{equation}
at $ s = s_{0} $, and 
\begin{equation}
\label{s-}
\psi_{1} (z) + \psi_{2} (z) = - \xi \left[ \varphi_{1} (z) + \varphi_{2} (z)\right]  , \; \varphi_{1} (z) - \varphi_{2} (z) = -\xi \left[ \psi_{1} (z) - \psi_{2} (z)\right]  
\end{equation}
at $ s = -s_{0} $.
The coefficient $ \xi $ in these expressions is defined below:
\begin{equation}
\label{xi}
\xi = \frac{\hbar k_{0}}{mc\left( s_{0} + 1 \right) } = \frac{\hbar c k_{0}}{\varepsilon_{0} + mc^{2}},
\end{equation}
where the definition is used $ \varepsilon_{0} = \sqrt{m^{2}c^{4} + (\hbar ck_{0})^2 }$.

Let us find the explicit form of the spinors which define eigen bispinors  (\ref{spinors_S}) of the spin operator  (\ref{S_e}). For this we introduce independent functions  $ F_{+} = \psi_{1} + \psi_{2} $ and $ \Phi_{+} = \varphi_{1} - \varphi_{2} $ for the state with $ s = s_{0} $, and functions  $ F_{-} = \psi_{1} - \psi_{2} $ and $ \Phi_{-} = \varphi_{1} + \varphi_{2} $ for the state with $ s = -s_{0} $. Using the quantum number  $ \sigma = \pm 1 $, which defines the eigenvalues  $ s = \sigma s_{0} = \sigma \varepsilon_{0}/mc^2 $,  
and expressing $ \psi_{1,2} $ and $ \varphi_{1,2} $ in terms of $ F_{\pm} $ and$ \Phi_{\pm} $ through the relations  (\ref{s+})-(\ref{s-}), we can represent the spinors in the following form: 
\begin{equation}
\label{VarPsi_+s}
\psi_{+} (z) = { e^{-i\phi_{0} /2 } \left( F_{+} + \xi \Phi_{+} \right)  \choose e^{i\phi_{0} /2 } \left( F_{+} - \xi \Phi_{+} \right) } , \quad  \varphi_{+} (z) = { e^{-i\phi_{0} /2 } \left( \xi F_{+}  + \Phi_{+} \right) \choose e^{i\phi_{0} /2 } \left( \xi F_{+} - \Phi_{+} \right) } ,
\end{equation}
\begin{equation}
\label{VarPsi_-s}
\psi_{-} (z) = { e^{-i\phi_{0} /2 } \left( F _{-}- \xi \Phi_{-} \right)  \choose - e^{i\phi_{0} /2 } \left( F_{-} + \xi \Phi_{-} \right) } , \quad  \varphi_{-} (z) = { - e^{-i\phi_{0} /2 } \left( \xi F_{-}  - \Phi_{-} \right) \choose e^{i\phi_{0} /2 } \left( \xi F_{-} + \Phi_{-} \right) } .
\end{equation}

Substituting now the obtained spinors (\ref{VarPsi_+s}) and (\ref{VarPsi_-s}) in DE (\ref{Eqs_2D}), we again obtain four equations for the two functions  $ F (z) $ and $ \Phi (z) $, which can be transformed to the next form
\begin{eqnarray}
-i\hbar c\frac{d F}{dz} + i\hbar c k_{\perp}\sigma  F - \hbar c k_{0}\xi \Phi - \left( E - V + mc^{2} \right)\Phi = 0 \, ,\nonumber \\
- i\hbar c\xi \frac{d F}{dz} + i\hbar c k_{\perp}\sigma  \xi F - \hbar c k_{0} \Phi - \left( E - V + mc^{2} \right) \xi \Phi = 0 \, , \nonumber \\
- i\hbar c\frac{d \Phi}{dz} - i\hbar c k_{\perp}\sigma  \Phi + \hbar c k_{0} \xi F - \left( E - V - mc^{2} \right) F = 0 \, , \nonumber \\
- i\hbar c\xi \frac{d \Phi}{dz} - i\hbar c k_{\perp}\sigma \Phi + \hbar c k_{0} F - \left( E - V + mc^{2} \right) \xi F = 0 \, . \nonumber 
\end{eqnarray}

Similarly to above, we see that the four equations can be reduced to two ones for the two unknown functions $F,\, \Phi  $: 
\begin{eqnarray}
\label{Eqs_fg_s}
-i\hbar c\frac{d F}{d z} + i\hbar ck_{\perp}\sigma  F = \left( E - V + \varepsilon_{0} \right) \Phi  \, , \nonumber \\
-i\hbar c\frac{d \Phi}{d z} - i\hbar ck_{\perp}\sigma  \Phi = \left( E - V - \varepsilon_{0} \right) F \, ,
\end{eqnarray}
where the Eq. (\ref{xi}) is used.

For the rectangular QW (\ref{V_123}) we obtain that functions $ F_{j} (z) $ $(j=L,C,R)$ are defined by Eq. (\ref{Eq_f}), and get the following relation for functions   $ \Phi_{j} (z) $:
\begin{equation}
\label{g_j_s}
\Phi_{j}(z) =  - i\frac{\hbar c}{ E - V_j + \varepsilon_{0} } \left( F' - \sigma k_{\perp} F \right). 
\end{equation}

\section{\label{DS-QW}Discrete spectrum of the quantum wall}
 Let us solve DE for electrons which are captured by the QW  (\ref{V_123}). In this case the function (\ref{Psi_2D}) can be written as
\begin{equation}
\label{Psi_123}
\Psi (x,y,z) = \left\lbrace \begin{array}{ll}
\Psi_{L} & \textrm{at  $ z< a $}\,  ,\\
\Psi_{C} & \textrm{at  $ a\leq z\leq b $}\\
\Psi_{R} & \textrm{at  $ z> b $}\, ,
\end{array} \right. , 
\end{equation}
where $ \Psi_{j}$ are the solutions of DE in the corresponding regions  $j=L,C,R $. At the boundaries  $ z=a $ and $ z=b $ the solutions should transform to each other continuously.  Therefore, the boundary conditions should be satisfied:
\begin{equation}
\label{bcond}
 \Psi_{L} (x,y,z=a) = \Psi_{C} (x,y,z=a)\, , \quad \Psi_{C} (x,y,z=b) = \Psi_{R} (x,y,z=b) \, .
\end{equation}

Let us count the energy from the value   $ V_{C} $. This means that we can set   $ V_{C} = 0 $. Then we have  $ E - V_{C} \longrightarrow E $ and $ V_{L,R} - V_{C} \longrightarrow V_{L,R} > 0 $ , where we have chosen  $V_{L}\ge V_{R} $ without loss of generality.  We are interested in the states with the energy $ E = \varepsilon_{\perp} + \mathcal{E} $, where $ \varepsilon_{\perp} $ is defined in  (\ref{vareps_2})). It is clear that the bound states have the energies  which satisfy the inequality $ 0 < \mathcal{E} < V_{R} $. The QW at $ V_{L} = V_{R} $ is symmetric, and at $ V_{L} \neq V_{R} $ it is asymmetric.  Below we will consider nonrelativistic energies and  assume strong inequality   $ V_{L} \ll mc^{2} $ to be valid when antiparticle sector is not important. It is known that at the given energy  the solutions, which correspond to antiparticles beyond  the central region, appear at rather large values of the potential. In this case the decreasing solutions take place at  $ V_{L,R} \geq mc^{2} $, and the solutions that correspond to real antiparticles, take place at  $ V_{L,R} \geq 2mc^{2} $. This is known as Klein paradox.

Different character of the solutions (\ref{varPsi_eps}), (\ref{VarPsi_+})-(\ref{VarPsi_-}) and (\ref{VarPsi_+s})-(\ref{VarPsi_-s}), which correspond to different spin states, results in different boundary conditions. It is worth to  mention that even the considered here limit of small energies is different from the standard nonrelativistic result, because each solution of DE is characterized by two functions, one of which satisfies Eq.  (\ref{Eq_f}), and the second function is determined by the first one and its derivative via the relations   (\ref{g_j_eps}), (\ref{g_j_mu}) or (\ref{g_j_s}). In the region $ C $  the right hand side of Eq. (\ref{Eq_f}) for the considered energies contains the constant multiplier $ E^{2} - \varepsilon_{\perp}^{2} > 0 $. Hence   (see (\ref{Eq_f})), function  $ f_{C}(z) $ should satisfy the equation  $ d^{2} f_{C}/dz^{2} = -q^{2} f_{C} $, the general solution of which is an oscillating function. Let us represent the latter in the form of  the linear combination of two exponential functions 
\begin{equation}
\label{f_C}
 f_{C}(z) = A_{C} e^{iqz} + B_{C} e^{-iqz} \, .
\end{equation}
The energy of the corresponding state is determined by the parameter  $ q $,
\begin{equation}
\label{E(q)}
E = \sqrt{\varepsilon_{\perp}^{2} +  \hbar^{2} c^{2}q^{2} } = \sqrt{m^{2} c^{4} +  \hbar^{2}c^{2}( k^{2} +  q^{2})} \equiv mc^{2} + \mathcal{E}(\mathbf{k},q) \, ,
\end{equation}
and represents a band in  $xy$-plane. 

Beyond the layer, in the regions $ z< a $  and   $ z> b $,  the negative coefficient in the right hand side of Eq.  (\ref{Eq_f}) corresponds to the energies of the bound states: 
$ E^{2} - \varepsilon_{\perp}^{2} < 0 $. Functions  $ f_{L,R} $ in these regions satisfy the equation  $ d^{2} f_{L,R}/dz^{2} = \kappa_{L,R}^{2} f_{L,R} $. Taking into account Eq. (\ref{E(q)}), we get 
\[ \left( E - V_{L,R} \right)^{2} - \varepsilon_{\perp}^{2} = -2EV_{L,R} \left( 1- \frac{(\hbar cq)^{2}}{2EV_{L,R}} - \frac{V_{L,R}}{2E} \right) = -\kappa_{L,R}^{2} \, . \]
In the considered case of non-relativistic energies, when the values $ V_{L}/mc^{2} \ll 1 $ and $ \mathcal{E}(\mathbf{k}, q)/mc^{2} \ll 1 $ can be neglected as comparing with 1,  we obtain $ E \approx mc^{2} $. In this approximation  nonrelativistic expression is valid for the parameter $ \kappa_{L,R} $: 
\begin{equation}
\label{kapa_LR}
\kappa_{L,R} = \sqrt{ \frac{2m }{\hbar^{2} }V_{L,R} \left( 1 - \frac{\hbar^{2} q^{2}}{2mV_{L,R}}\right) } = \sqrt{(\kappa^{(0)}_{L,R})^{2} - q^{2}} ,  
\end{equation}
where
\begin{equation}
\label{kappa_0}
\kappa^{(0)}_{L,R} = \sqrt{\frac{2m }{\hbar^{2} }V_{L,R}} \, .
\end{equation}
Therefore, as it is expected, the formal solutions in these regions are exponentially increasing and decreasing functions. Since only the decreasing solution has a physical meaning, we conclude that in the left region the solution is given by the function
\begin{equation}
\label{f_L}
f_{L}(z) = C_{L} e^{\kappa_{L} (z-a)} \, , 
\end{equation}
which decreases at  $ z \rightarrow -\infty $. The solution in the right region, 
\begin{equation}
\label{f_R}
f_{R}(z) = C_{R} e^{-\kappa_{R} (z-b)} \, , 
\end{equation}
 vanishes at  $ z \rightarrow \infty $. 

In all three cases function $ f_j(z) $ is given by one of the functions with the given electric spin polarization, while the second function is given by the expression (\ref{g_j_eps}). The function,  which determines the states with the given magnetic spin polarization, is obtained by the substitution  $ f_j(z) \rightarrow \psi_j (z) $, and function $ \varphi_j (z) $ is determined by the expression (\ref{g_j_mu}). For the states with given spin polarization  $ \hat{\mathcal{S}}_{0} $ (see (\ref{S_e})) the eigenfunctionis $ f_j (z) \rightarrow F_j  (z) $, and function $ \Phi_j  (z) $ is expressed via connected with the former one using the relation  (\ref{g_j_s}). Taking this into account, we find the following results for functions (\ref{g_j_eps}),  (\ref{g_j_mu}) and  (\ref{g_j_s}), respectively:
\begin{eqnarray}
\label{g_CLR}  
g_{C}(z) = \frac{\hbar c}{E + mc^{2}} \left[ \left( q + i \sigma k \right) A_{C} e^{iqz} - \left( q - i \sigma k \right) B_{C} e^{-iqz} \right] \, , \nonumber \\
g_{L}(z) = -i\frac{\hbar c \left(\kappa_{L} - \sigma k \right) }{E + mc^{2} - V_{L}} C_{L} e^{\kappa_{L} (z-a)} \, , \\
g_{R}(z) = i\frac{\hbar c \left(\kappa_{R} + \sigma k \right) }{E + mc^{2} - V_{R}} C_{R} e^{-\kappa_{R} (z-b)} \, ;\nonumber 
\end{eqnarray}
\begin{eqnarray}
\label{varfi_CLR}
\varphi_{C}(z) = \sigma \frac{\hbar cq}{E + \varepsilon_{\perp}} \left( A_{C} e^{iqz} - B_{C} e^{-iqz} \right) \, , \nonumber \\
\varphi_{L} (z) = -i \sigma \frac{\hbar c\kappa_{L}}{E + \varepsilon_{\perp} - V_{L}} C_{L} e^{\kappa_{L} (z-a)} \, , \\
\varphi_{R} (z) = i \sigma\frac{ \hbar c\kappa_{R}}{E + \varepsilon_{\perp} - V_{R}} C_{R} e^{-\kappa_{R} (z-b)} \, ; \nonumber 
\end{eqnarray}
\begin{eqnarray}
\label{Fi_CLR}
 \Phi_{C}(z) = \frac{\hbar c}{E + \varepsilon_{0}} \left[ \left( q + i \sigma k_{\perp} \right) A_{C} e^{iqz} - \left( q - i \sigma k_{\perp} \right) B_{C} e^{-iqz} \right] \, , 
 \nonumber \\
\Phi_{L}(z) = -i\frac{\hbar c\left(\kappa_{L} - \sigma k_{\perp} \right) }{E + \varepsilon_{0} - V_{L}} C_{L} e^{\kappa_{L} (z-a)} \, ,  \\ 
\Phi_{R}(z) = i\frac{\hbar c\left(\kappa_{R} + \sigma k_{\perp} \right) }{E + \varepsilon_{0} - V_{R}} C_{R} e^{-\kappa_{R} (z-b)} \, .\nonumber
\end{eqnarray}
Here $ k_{\perp} $ is defined in (\ref{p_e_ort}) and $ \varepsilon_{0} $  -- in (\ref{xi}). It is easy to see that the found above sets of functions corresponding to different spin states, do not coincide.  

The coefficients $ A_{C} $, $ B_{C} $, $ C_{L} $ and $ C_{R} $, as well as the allowed values of the parameter   $ q $, are determined by the normalization condition and boundary conditions  (\ref{bcond}) required for the continuity of the wavefunction at the corresponding boundaries. The equality of the bispinors (\ref{bcond}) corresponds to the equality of the four components and is reduced to the equality of the two incoming functions.   Equality of functions which satisfy Eq. (\ref{Eq_f}), gives the following relations:
\[ C_{L} = A_{C} e^{iqa} + B_{C} e^{-iqa} \, , \quad  C_{R} = A_{C} e^{iqb} + B_{C} e^{-iqb} \, , \]
from which we can express the coefficients  $ C_{L} $ and $ C_{R} $ via the coefficients $ A_{C} $ and $ B_{C} $. Substituting this result into the condition of the equality of functions  (\ref{g_CLR}), (\ref{varfi_CLR}) or (\ref{Fi_CLR}), we get the system of two homogenous equations for the coefficients  $ A_{C} $ and $ B_{C} $:
\begin{eqnarray}
\label{eqsAB}
F_{L}^{\ast} e^{iqa} A_{C} + F_{L} e^{-iqa} B_{C} = 0 \, , \nonumber \\
F_{R} e^{iqb} A_{C} + F_{R}^{\ast} e^{-iqb} B_{C} = 0 \, .
\end{eqnarray}
Here the complex coefficients can be written as    $ F_{L,R} = \vert F_{L,R} \vert \exp \left(i\theta_{L,R} \right) $. Below we give the expressions for the coefficients for principally different physical situations:  \\
 \emph{i) the states with the given electric spin polarization } (see (\ref{eps}))
\begin{eqnarray}
\label{F_LR_eps}
F_{L} = \left( \kappa_{L} - \sigma \nu_{L} k \right) + i\left( 1 - \nu_{L} \right)q  \, , \nonumber \\
|F_{L}| = \sqrt{\left( \kappa_{L} - \sigma \nu_{L} k \right)^{2} + \left( 1 - \nu_{L} \right)^{2}q^{2}}\, ,\quad \tan \theta_{L} = \frac{\left( 1 - \nu_{L} \right)q}{\kappa_{L} - \sigma \nu_{L} k} \, ,\nonumber \\
F_{R} = \left( \kappa_{R} +\sigma \nu_{R} k \right) + i\left( 1 -\nu_{R} \right)q \, ,\\ 
|F_{R}| = \sqrt{\left( \kappa_{R} + \sigma \nu_{R} k \right)^{2} + \left( 1 - \nu_{R} \right)^{2}q^{2}}\, ,\quad \tan \theta_{R} = \frac{\left( 1 - \nu_{R} \right)q}{\kappa_{R} + \sigma \nu_{R} k} \, ; \nonumber 
\end{eqnarray}
\emph{ii) the states with the given magnetic spin polarization  } (see (\ref{mu}))
\begin{eqnarray}
\label{F_LR_mu}
F_{L} = \kappa_{L} + i\left( 1 - \nu_{L} \right)q  \, , \quad  F_{R} = \kappa_{R} + i\left( 1 - \nu_{R} \right)q \, , \nonumber \\
|F_{L}| = \sqrt{\kappa_{L}^{2} + \left( 1 - \nu_{L} \right)^{2}q^{2} } \, ,\quad \tan \theta_{L} = \frac{\left( 1 - \nu_{L} \right)q}{\kappa_{L} } \, , \\
|F_{R}| = \sqrt{\kappa_{R}^{2} +\left( 1 - \nu_{R} \right)^{2} q^{2}}\, ,\quad \tan \theta_{R} = \frac{\left( 1 - \nu_{R} \right)q}{\kappa_{R} } \, ; \nonumber 
\end{eqnarray}
\emph{iii) the states with the given  spin polarization $ \hat{\mathcal{S}}_{0} $ } (see (\ref{S_e}))
\begin{eqnarray}
\label{F_LR_se}
F_{L} = \left( \kappa_{L} - \sigma \nu_{L} k_{\perp} \right) + i\left( 1 - \nu_{L} \right)q  \, , \nonumber \\
|F_{L}| = \sqrt{\left( \kappa_{L} - \sigma \nu_{L} k_{\perp} \right)^{2} +\left( 1 - \nu_{L} \right)^{2} q^{2}}\, ,\quad \tan \theta_{L} = \frac{\left( 1 - \nu_{L} \right)q}{\kappa_{L} - \sigma \nu_{L} k_{\perp}} \, ,\nonumber \\
F_{R} = \left( \kappa_{R} +\sigma \nu_{R} k_{\perp} \right) + i\left( 1 -\nu_{R} \right)q \, ,\\ 
|F_{R}| = \sqrt{\left( \kappa_{R} + \sigma \nu_{R} k_{\perp} \right)^{2} + \left( 1 - \nu_{R} \right)^{2}q^{2}}\, ,\quad \tan \theta_{R} = \frac{\left( 1 - \nu_{R} \right)q}{\kappa_{R} + \sigma \nu_{R} k_{\perp}} \, . \nonumber 
\end{eqnarray}
Here the following notations are introduced:  $ \nu_{L,R} = V_{L,R}/(E+mc^{2}) $ in (\ref{F_LR_eps}), $ \nu_{L,R} = V_{L,R}/(E+\varepsilon_{\perp}) $ in (\ref{F_LR_mu}) and $ \nu_{L,R} = V_{L,R}/(E+\varepsilon_{0}) $ in (\ref{F_LR_se}). For nonrelativistic energies they take the values $ \nu_{L,R} \approx V_{L,R}/(2mc^{2}) $.

The condition of the existence of a non-trivial solution of Eqs.  (\ref{eqsAB}), 
\[ F_{L}^{\ast} F_{R}^{\ast} e^{-iq(b-a)} - F_{L} F_{R} e^{iq(b-a)} = 0 , \]
leads to the equation $ \sin \left( qd + \phi_{L} + \phi_{R} \right) = 0 $, which is equivalent to the condition  $ qd + \theta_{L} + \theta_{R} = \pi n $. Taking into account expressions (\ref{F_LR_eps})-(\ref{F_LR_se}), the latter condition can be rewritten in the form 
\begin{equation}
\label{cond1}
qd + \arcsin \frac{\left( 1 - \nu_{L} \right)q}{|F_{L}|} + \arcsin \frac{\left( 1 - \nu_{R} \right)q}{|F_{R}|} = \pi n  \, , \quad n = 1, 2, \ldots 
\end{equation}
 
\section{\label{q2d-s}Spectra of quasi-two-dimensional electrons with different spin states }

The roots of the transcendent Eq. (\ref{cond1}), $ q_{n} $, determine the allowed discrete values of the parameter   $ q $, and, hence, the energies of the bound states  (\ref{E(q)}) at $ q = q_{n} $, increasing with  $ n $ increasing. For nonrelativistic energies at $ \mathcal{E}( \mathbf{k}, q) \ll  mc^{2} $, the values $ \nu_{L,R} \ll 1 $ can be also neglected and the energy spectrum of 2D electron bands is
\begin{equation}
\label{calE}
\mathcal{E}_{n \sigma}(\mathbf{k}) =  \frac{\hbar^{2} \mathbf{k}^{2}}{2m} + \frac{\hbar^{2}q_{n \sigma}^{2}}{2m}\, .
\end{equation}

Although the energy spectrum  (\ref{calE}) looks as having a standard form of a double-degenerate spectrum with respect to the spin number  $ \sigma = \pm 1 $, it is not so, because, as we have shown above, the spin states of a relativistic particle are different  (see (\ref{F_LR_eps}), (\ref{F_LR_mu}), (\ref{F_LR_se})). Below we will show that the roots $q_{n \sigma}$ for the states with $\sigma =1$ and $\sigma =-1$ are different in each of the considered cases. The only exception is the case of the states with the given magnetic spin polarization  (\ref{F_LR_mu}), when the spectrum is indeed degenerate. Indeed, using expressions (\ref{F_LR_mu}) and taking into account (\ref{kapa_LR}) in Eq. (\ref{cond1}), we get the transcendent equation 
\[qd + \arcsin \frac{q}{\kappa^{(0)}_{L}} + \arcsin \frac{q}{\kappa^{(0)}_{R}} = \pi n  \, ,\]
with the values  $ \kappa^{(0)}_{L,R} $ defined in Eq. (\ref{kappa_0}), which coincides with  the known equation of Quantum Mechanics for a  nonrelativistic particle in a  1D rectangular potential well (see, e.g.,  \cite{Davydov}), and has the roots that are independent of  $\sigma $.  

Therefore, the subbands of electrons with the given magnetic spin polarization, (\ref{calE}), are degenerate with respect to the spin states and are determined by the 'initial' dispersion in which the presence of the external potential is reflected in spatial dimensional quantization, only, without any manifestation of SOI, and, hence, without Rashba interaction. 

Let us now consider the states with the given electric spin polarization  (\ref{F_LR_eps}). Substituting expression  (\ref{F_LR_eps}) into condition  (\ref{cond1}), we get the equation
\begin{equation}
\label{cond1m}
 qd + \arctan \frac{q}{\kappa_{L} - \sigma \nu_{L} k} + \arctan \frac{q}{\kappa_{R} + \sigma \nu_{R} k} = \pi n \, .
\end{equation}
It is easy to see that the roots of this equation depend  not only on the modulus of the wave number $ \bm
 {k} $, but also on the spin number 
 $ \sigma $. The above equation contains these two values in the combination of their product $ \sigma k $,  only, and, hence,  the roots of the equation depend on them as  $ q_{n \sigma}(\mathbf{k}) = q_{n}(\sigma k) $. Taking into account that   $ \sigma = \pm 1 $ and  $ \sigma^{2} = 1 $, we can represent the roots of Eq.  (\ref{cond1}) as the sum of odd and even components:   $ q_{n}(\sigma k) = q^{(\mathrm{ev})}_{n}(k) + \sigma q^{(\mathrm{od})}_{n}(k) $. Here the expansions   $ q^{(\mathrm{ev})}_{n}(k) $ and  $ q^{(\mathrm{od})}_{n}(k) $ with respect to $ k $ contain even and odd powers of $ k $, respectively. For small values of $ k $ or assuming    $ q^{(\mathrm{ev})}_{n}(k) \approx q^{(\mathrm{ev})}_{n}(0) + (1/2) \ddot{q}^{(\mathrm{ev})}_{n}(0) k^{2} $ and  $ q^{(\mathrm{od})}_{n}(k) \approx \dot{q}^{(\mathrm{od})}_{n}(0) k $, we get the expression
\[
q^{2}_{n}(\sigma k) \simeq (q^{(\mathrm{ev})}_{n}(0))^2 + \left[ q^{(\mathrm{ev})}_{n}(0) \ddot{q}^{(\mathrm{ev})}_{n}(0) + \left( \dot{q}^{(\mathrm{od})}_{n}( 0 )\right)^{2} \right] k^{2} + 2\sigma q^{(\mathrm{ev})}_{n}(0) \dot{q}^{(\mathrm{od})}_{n}(0) k \, ,
\]
where dots denote the derivatives of the corresponding functions. The above expression determines the change of the dispersion of 2D electrons, (\ref{calE}), depending on their spin state and   quasimomentum  modulus $ k $. 
 
Therefore, for the states with the given electric spin polarization we obtain isotropic in the $ \mathbf{k} $-space Rashba dispersion law with the spin degenerate 2D electron subbands:
\begin{equation}
\label{calE_eps}
\mathcal{E}_{n\sigma}(\mathbf{k}) = \mathcal{E}_{n}(0) + \frac{\hbar^{2} \mathbf{k}^{2}}{2m^{\ast}_{n}} + \sigma \alpha_{n} k \, ,
\end{equation}
where  
\begin{equation}
\label{E-m-alf}
 \mathcal{E}_{n}(0) = \frac{\hbar^{2}\left(  q^{(\mathrm{ev})}_{n}(0)\right)^{2} }{2m} \,  
 \end{equation}
 is the energy of the bottom of the corresponding subband, 
 \begin{equation}
  m^{\ast}_{n} = \frac{m}{1 + q^{(\mathrm{ev})}_{n}(0) \ddot{q}^{(\mathrm{ev})}_{n}(0) + \left( \dot{q}^{(\mathrm{od})}_{n}(0)\right)^{2} } \, 
 \end{equation}
 is the corresponding renormalized mass, and 
 \begin{equation} 
 \alpha_{n} = \frac{\hbar^{2} }{m} q^{(\mathrm{ev})}_{n}(0) \dot{q}^{(\mathrm{od})}_{n}(0) \,  
\end{equation}
is the parameter, which can be identified with Rashba SOI parameter. 
 
Finally, let us consider the states with the given spin polarization  $ \hat{\mathcal{S}}_{\mathbf{e}_0} $. We can set for definiteness $ \phi_{0} = \pi /2 $. Then $ \mathbf{e}_0 = \mathbf{e}_{y} $, $ \mathbf{e}_{\perp} = \mathbf{e}_{x} $ and, respectively, $ k_{0} = k_{y} $,  $ k_{\perp} = k_{x} $. In this case we again obtain from Eq.  (\ref{calE}) spin-splitted energy dispersion law, which is now anisotropic:  
\begin{equation}
\label{calE_s}
\mathcal{E}_{n\sigma}(\mathbf{k}) = \mathcal{E}_{n}(0) + \frac{\hbar^{2} k_{y}^{2}}{2m} + \frac{\hbar^{2} k_{x}^{2}}{2m^{\ast}_{n}} + \sigma \alpha_{n} k_{x} \, .
\end{equation}
It is easy to see that its explicit form is different from  (\ref{E-m-alf}) because in this case Eq.   (\ref{cond1m}) contains not the modulus of the wavevector, but its projection, which means the anisotropy of the $ xy $-plane. The meaning of notations in  (\ref{calE_s}) is similar to the previous case.  

To conclude, we note that the spin degree of freedom is characterized by the two quantum numbers  $ \sigma =1 $ and $ \sigma = -1 $, which can have different physical meaning. The impact of the external inhomogenous potential (the boundary of the QW in the considered above  case) on the electron energy depends on the state, to which the given spin number corresponds. The potential, created by the electric field, does not affect the states with the given magnetic spin polarizations, while it leads to the splitting of the  2D electron bands with respect to spin and modifies the dispersion laws in the cases of the electric spin polarization or polarization $ \hat{\mathcal{S}}_{\mathbf{e}_0} $. This means the appearance of SOI in the latter  cases. 

Worth mentioning that the splitting of the bands with different spin numbers takes place in asymmetric potential wells only. In symmetric wells, $ V_L=V_R $, the roots of Eq.  (\ref{cond1}) don't depend on  $ \sigma $, and, taking into account the relation $ \arctan x + \arctan y = \arctan \left[ ( x+y )/(1-xy)\right]$, we see from the condition (\ref{cond1m}) that Rashba parameter equals zero, as it should be.    Nevertheless, SOI is non-zero also in symmetric potentials and manifests itself in the second order through the renormalization of 2D electron mass.  Qualitatively similar results are obtained in   \cite{Bernardes}, where it is shown for two 2D electron bands that SOI results in the second order in the change of dispersion laws in a symmetric QW.   

\section{\label{WF-dS} Wavefunctions of electrons with different spin states }

We have obtained above the eigen wavefunctions of different spin operators which have to correspond to eigen wavefunctions of DE. The latter functions allow calculating the probabilities of physical observables and probabilities of the results of various measurements. These probabilities are given by  expressions which are bilinear with respect to  $ \Psi $ and $ \Psi^{\ast} $. For instance, the probability density of particle presence  at   $ \mathbf{r} $ has a standard form  $ \rho (\mathbf{r}) = \Psi^{\dagger} \Psi $, the current density probability is $ \mathbf{j} = c \Psi^{\dagger} \bm{\hat{\alpha}} \Psi $, the probability density of the eigen spin momentum which is proportional to the vector,  is $ \mathbf{S} = \Psi^{\dagger} \bm{\hat{\Sigma}} \Psi $, and so on. To calculate these values, we find below the explicit form of the wavefunctions. 

Let us set $ a = z_{0} - d/2 $ and $ b = z_{0} + d/2 $, where $ z_{0} $  is the center of the potential layer, and  $ d $ is its width. Taking into account the relation (\ref{cond1}), we get $ qa = qz_{0} + (\theta_{L} + \theta_{R})/2 - \pi n/2 $ and $ qb = qz_{0} - (\theta_{L} + \theta_{R})/2 + \pi n/2 $. Choosing the origin of coordinates at the center of the QW,   $ z_{0} = 0 $, the solution of Eqs. (\ref{eqsAB}) has the form 
\begin{eqnarray}
\label{ABA_LR}
A_{C} = - \frac{i}{2} A_{n} e^{i\left( \Delta \phi + \pi n/2 \right) } \, , \quad 
B_{C} = \frac{i}{2} A_{n} e^{-i\left( \Delta \phi + \pi n/2 \right) } \, ,\nonumber \\
C_{L} = A_{n} \sin \theta_{L} \, , \qquad C_{R} = (-1)^{n-1} A_{n} \sin \theta_{R} \, ,
\end{eqnarray}
where $ \Delta \phi = \left( \theta_{L} - \theta_{R}\right) /2 $, and $ A_{n} $ are normalization constants for the given value of  $n$. The function  $ f(z) $ in (\ref{f_C}) in the corresponding regions can be given by expressions: 
\begin{eqnarray}
\label{f_sol}
f_{L}(z) = A_{n} \sin \theta_{L} e^{\kappa_{L} (z+d/2)} \, ,
\nonumber \\
f_{C}(z) = A_{n} \sin \left( q_{n}z + \Delta \phi + \frac{\pi}{2} n  \right) \, , \\
f_{R}(z) = (-1)^{n-1} A_{n} \sin \theta_{R} e^{-\kappa_{R} (z-d/2)} \, , 
\nonumber 
\end{eqnarray}
where phases $ \theta_{L,R} $ depend on the given spin state  $ f(z) $. The same is valid for functions   $ \psi(z) $ and $ F (z) $, respectively.

Using these functions, we can calculate the corresponding bispinors for all three possible states:\\
\emph{i) states with the given electric spin polarization   }  
\begin{eqnarray}
\label{g_sol}
g_{C}(z) = -i\frac{\hbar c}{E + mc^{2}} A_{n} \left[ q_{n} \cos \left( q_{n}z + \Delta \phi + \frac{\pi}{2} n \right) - \sigma k \sin \left( q_{n}z + \Delta \phi + \frac{\pi}{2} n \right)  \right]  \, , \nonumber \\
g_{L}(z) = -i\frac{\hbar c\left(\kappa_{L} - \sigma k \right) }{E + mc^{2} - V_{L}} A_{n} \sin \theta_{L} e^{\kappa_{L} (z+d/2)} \, , \\
g_{R}(z) = (-1)^{n-1} i\frac{\hbar c\left(\kappa_{R} + \sigma k \right) }{E + mc^{2} - V_{R}} A_{n} \sin \theta_{R} e^{-\kappa_{R} (z-d/2)} \, ; \nonumber
\end{eqnarray}
\emph{ii) states with the  given magnetic spin polarization } 
\begin{eqnarray}
\label{varfi_sol}
\varphi_{C}(z) = -i\frac{\sigma \hbar cq_{n}}{E + \varepsilon_{\perp}} A_{n} \cos \left( q_{n}z + \Delta \phi + \frac{\pi}{2} n  \right) \, , \nonumber \\
\varphi_{L} (z) = -i \frac{\sigma \hbar c\kappa_{L}}{E + \varepsilon_{\perp} - V_{L}} A_{n} \sin \theta_{L} e^{\kappa_{L} (z+d/2)} \, , \\
\varphi_{R} (z) = (-1)^{n-1} i\frac{\sigma \hbar c\kappa_{R}}{E + \varepsilon_{\perp} - V_{R}}  A_{n} \sin \theta_{R} e^{-\kappa_{R} (z-d/2)} \, ; \nonumber
\end{eqnarray}
\emph{iii) states with the  given polarization $ \hat{\mathcal{S}} $ }
\begin{eqnarray}
\label{Phi_sol}
\Phi_{C}(z) = -i\frac{\hbar c}{E + \varepsilon_{1}} A_{n} \left[ q_{n\sigma} \cos \left( q_{n\sigma}z + \Delta \phi + \frac{\pi}{2} n \right) - \sigma k_{\perp} \sin \left( q_{n\sigma}z + \Delta \phi + \frac{\pi}{2} n \right)  \right]  \, , \nonumber \\
\Phi_{L}(z) = -i\frac{\hbar c\left(\kappa_{L} - \sigma k_{\perp} \right) }{E + \varepsilon_{0} - V_{L}} A_{n} \sin \theta_{L} e^{\kappa_{L} (z+d/2)} \, , \\
\Phi_{R}(z) = (-1)^{n-1} i\frac{\hbar c\left(\kappa_{R} + \sigma k_{\perp} \right) }{E + \varepsilon_{0} - V_{R}} A_{n} \sin \theta_{R} e^{-\kappa_{R} (z-d/2)} \, . \nonumber
\end{eqnarray}

Using these expressions, we can calculate the probability density of electron distribution in the QW depth. We will consider below the ground electron state  $ n = 1 $. We get the following results for the three cases:\\
\emph{ states with the  the given electric spin polarization } 
\[\rho^{(\epsilon)} _{\sigma } (z) = \mid f_{\sigma} \mid^{2} + \mid g_{\sigma} \mid^{2} = \]
\begin{equation}
\label{rho_eps}
 = \frac{\mid A^{(\epsilon)}_{1} \mid^{2}}{2} \left[ C_{1} + C_{2} \cos \left( 2q_{1\sigma}z + \theta_{L} - \theta_{R} \right) + \sigma \frac{\hbar^{2}kq_{1\sigma}}{2m^{2}c^{2}} \sin \left( 2q_{1\sigma}z + \theta_{L} - \theta_{R} \right) \right] \, ;
\end{equation}
 \emph{states with the  the given magnetic spin polarization } 
\[\rho^{(\mu)}_{\sigma } (z) = \left( 1+\lambda^{2} \right) \left( \mid \psi_{\sigma} \mid^{2} + \mid \varphi_{\sigma} \mid^{2} \right) = \]
\begin{equation}
\label{rho_mu}
= \frac{\mid A^{(\mu)}_{1} \mid^{2}}{2} \left[ C_{1} + C_{2} \cos \left( 2q_{1}z + \theta_{L} - \theta_{R} \right) \right] \, ;
\end{equation}
  \emph{states with the  the given polarization}  $ \hat{\bm{\mathcal{S}}} $ 
\[\rho^{(\mathcal{S})}_{\sigma } (z) = 2 \left( 1+\xi^{2} \right) \left( \vert F_{\sigma} \vert^{2} + \mid \Phi_{\sigma} \mid^{2} \right) = \]
\begin{equation}
\label{rho_s}
= \frac{\vert A^{(\mathcal{S})}_{1} \vert^{2}}{2} \left[ C_{1} + C_{2} \cos \left( 2q_{1\sigma}z + \theta_{L} - \theta_{R} \right) + \sigma \frac{\hbar^{2}k_{\perp}q_{1\sigma}}{2m^{2}c^{2}} \sin \left( 2q_{1\sigma}z + \theta_{L} - \theta_{R} \right) \right] \, .
\end{equation}
The coefficients  $ \left( 1+\lambda^{2} \right) $ in (\ref{rho_mu}) and $ 2 \left( 1+\xi^{2} \right) $ in  (\ref{rho_s}) are included in normalization constants, and coefficients $ C_{1} $ and $ C_{2} $ are defined as  
\[C_{1} = 1 + \frac{\hbar^{2} \left(  k^{2}  +q_{1\sigma}^{2}\right) }{4m^{2}c^{2}} \, , \quad 
C_{2} = 1 + \frac{\hbar^{2} \left(k^{2}- q_{1\sigma}^{2}   \right) }{4m^{2}c^{2}} \]
in (\ref{rho_eps}), as 
\[C_{1} = 1 + \frac{\hbar^{2} q_{1}^{2} }{4m^{2}c^{2}} \, , \quad C_{2} = 1 - \frac{\hbar^{2} q_{1}^{2} }{4m^{2}c^{2}} \]
in (\ref{rho_mu}) and as 
\[C_{1} = 1 + \frac{\hbar^{2} \left(k_{\perp}^{2}+ q_{1\sigma}^{2}   \right) }{4m^{2}c^{2}} \, , \quad 
C_{2} = 1 + \frac{\hbar^{2} \left(  k_{\perp}^{2}-q_{1\sigma}^{2} \right) }{4m^{2}c^{2}} \]
in (\ref{rho_s}). For the nonrelativistic energies these coefficients are close to unity,   $ C_{1} \simeq C_{2} \simeq 1 $.

It follows from above that SOI does not affect the probability density of electron distribution inside the QW in the case of electrons with the given magnetic spin polarization.  Distribution probability density of electrons with the given electric spin polarization or with the given polarization   $ \hat{\mathcal{S}} $ do depend on the spin number. In particular, electrons with opposite spin numbers are shifted towards the opposite boundaries of the QW.   

The probability densities  of the vector spin operators  (\ref{mu}), (\ref{eps}) and (\ref{genS}) determine the vectors which characterize polarization properties of the corresponding spin states. Electric spin polarization (\ref{eps}) is non-zero for the states with the given electric spin polarization, only, and has   $ z $-component, only: $ \langle \bm{\hat{\epsilon}} \rangle^{(\epsilon )}_{\sigma } =  (\Psi^{(\epsilon )}_{\sigma })^{\dagger} \bm{\hat{\epsilon}} \Psi^{(\epsilon )}_{\sigma } = \sigma p \rho^{(\epsilon )}_{\sigma } \mathbf{e}_{z} $. For the states with the   given magnetic spin polarization $ \bm{\mu } $ and with the polarization  $ \hat{\bm{\mathcal{S}}} $ we have $ \langle \hat{\epsilon}_{z} \rangle^{(\mu)}_{\sigma } = \langle \hat{\epsilon}_{z} \rangle^{(\mathcal{S})}_{\sigma } = 0 $.

Polarization of the electron spin can be characterized by the mean values of vectors  $ \langle \bm{\hat{\mu}} \rangle $ and $ \langle \bm{\hat{\mathcal{S}}} \rangle $. In the nonrelativistic case when we can neglect terms that are proportional to   $ 1/c^{2} $, these vectors are reduced to the mean values  $ \langle \bm{\hat{\mu}} \rangle \approx \langle \bm{\hat{\Sigma}} \rangle $ and $ \langle \bm{\hat{\mathcal{S}}} \rangle \approx \langle \bm{\hat{\Omega}} \rangle $, respectively. Let us write down these vectors for the considered above three different spin states: \\
\emph{i)  with the given electric spin polarization }
\begin{eqnarray}
\label{SiOM_eps}
\langle \bm{\hat{\Sigma}} \rangle^{(\epsilon)}_{\sigma } = \sigma \left( \mid f \mid^{2} - \mid g \mid^{2} \right) \left( -\mathbf{e}_{x}\sin \phi + \mathbf{e}_{y} \cos \phi \right) = \sigma \left( \mid f \mid^{2} - \mid g \mid^{2} \right)\mathbf{e}_{z}\times \mathbf{e}_{\mathbf{k}} \, , \nonumber \\
\langle \bm{\hat{\Omega}} \rangle^{(\epsilon)}_{\sigma } = \sigma \left( \mid f \mid^{2} + \mid g \mid^{2} \right) \mathbf{e}_{z}\times \mathbf{e}_{\mathbf{k}} = \sigma \rho^{(\epsilon)}_{\sigma } \mathbf{e}_{z}\times \mathbf{e}_{\mathbf{k}} \, ;
\end{eqnarray}
\emph{ii)  with the given magnetic spin polarization }
\begin{eqnarray}
\label{SiOM_mu}
\langle \bm{\hat{\Sigma}} \rangle^{(\mu)}_{\sigma } = \sigma \left[ \left( 1 - \lambda^{2} \right)  \left( \mid \psi \mid^{2} + \mid \varphi \mid^{2} \right) \mathbf{e}_{z} +  \mathcal{O}(c^{-2}) \mathbf{e}_{z}\times \mathbf{e}_{\mathbf{k}} \right]  \, , \nonumber \\
\langle \bm{\hat{\Omega}} \rangle^{(\mu)}_{\sigma } = \sigma \left[ \left( 1 + \lambda^{2} \right)  \left( \mid \psi \mid^{2} - \mid \varphi \mid^{2} \right) \mathbf{e}_{z} +  \mathcal{O}(c^{-2}) \mathbf{e}_{\mathbf{k}} \right] \, ;
\end{eqnarray}
\emph{iii)  with the given polarization $ \hat{\bm{\mathcal{S}}} $ }
\begin{eqnarray}
\label{SiOM_s}
\langle \bm{\hat{\Sigma}} \rangle^{(\mathcal{S})}_{\sigma } = 2\sigma  \left( 1 + \xi^{2} \right) \left( \mid F \mid^{2} - \mid \Phi \mid^{2} \right) \mathbf{e}_0 \, , \nonumber \\
\langle \bm{\hat{\Omega}} \rangle^{(\mathcal{S})}_{\sigma } \simeq 2\sigma  \left( 1 - \xi^{2} \right) \left( \mid F \mid^{2} + \mid \Phi \mid^{2} \right) \mathbf{e}_0 \, .
\end{eqnarray}
In Eqs. (\ref{SiOM_eps})-(\ref{SiOM_s}) the vector $ \mathbf{e}_{\mathbf{k}} $  is defined as a unit vector in the direction of the wavevector, the vector  $ \mathbf{e}_0 $ as a unit vector in the chosen  direction of the   vector $ {\hat{\bm{\mathcal{S}}}} $ quantization. 

We remind that in all cases the bispinors are characterized by the two functions, one of which is a small component (functions $ g $, $ \varphi $ and $ \Phi $). In the nonrelativistic approximation this small component can be neglected in expressions  (\ref{SiOM_eps}), (\ref{SiOM_mu}) and (\ref{SiOM_s}). In this case $ \left\langle \bm{\hat{\Sigma}} \right\rangle = \left\langle \bm{\hat{\Omega}} \right\rangle $, and, hence, in a nonrelativistic case the spin polarization is characterized by one vector  $ \mathbf{S} = \left\langle \bm{\hat{\Sigma}} \right\rangle $, only.

\section{\label{Concl}Concluding remarks}

Let us compare the obtained above results for DE in the nonrelativistic limit with the results which follow from Schr\"{o}dinger equation with account of SOI: 
\begin{equation} 
\label{SEq}
\left[ -\frac{\hbar^{2}}{2m}\triangle + V(\mathbf{r}) - i \frac{\hbar^{2}}{4m^{2}c^{2}}  \left(\hat{\bm{\sigma}} \times \bm{\nabla} V \right)\cdot \bm{\nabla} \right]  \Psi = E \Psi .
\end{equation}
The last term in  (\ref{SEq}), which binds electron spin with its momentum is traditionally called SOI. It is obtained in the result of the expansion of DE with respect to $ 1/c $ with the accuracy of the second order  \cite{Bete, LandauIV, Davydov}. In the case when the external potential varies along one direction, e.g., $ z$-axis, and is homogenous in a perpendicular $ xy$-plane, the states of a particle with the given value of its 2D momentum is described by the function (\ref{Psi_2D}), in which $ \varPsi (z) $ is a spinor. Substituting  (\ref{Psi_2D}) into Schr\"{o}dinger equation, we obtain the following equation for the spinor   
\[
\left[ - \frac{\hbar^{2}}{2m} \frac{d^{2}}{d z^{2}} + \frac{\hbar^{2} k^{2} }{2m} + V(z) + \frac{\hbar^{2}}{4m^{2}c^{2}} \frac{dV}{dz} k \hat{\Lambda}_R  \right] \varPsi = E \varPsi \, ,
\]
which contains unknown space-independent matrix $  k \hat{\Lambda}_R $ (see Eq. (\ref{Lambda})). Therefore, the eigen wavefunction has to be proportional to the eigen spinors of this matrix,   $ \chi_{\sigma} $, which correspond to eigenvalues  $ \sigma = \pm 1 $, that are defined in Eq. (\ref{spinors}). We see that the solution takes the form  $ \varPsi (z) = f_{\sigma} (z) \chi_{\sigma} $. In the case of a rectangular QW Â (\ref{V_123}) we get $ dV/dz = \left( V_{C} - V_{L} \right)  \delta (z-a) + \left( V_{R} - V_{C} \right) \delta (z-b) $, and, hence, the wavefunction $ f_{\sigma} (z) $  satisfies the equation 
\[ \left\lbrace - \frac{\hbar^{2}}{2m} \frac{d^{2}}{d z^{2}} + V(z) - \sigma \frac{\hbar^{2}k}{4m^{2}c^{2}} \left[ V_{L} \delta (z-a) - V_{R} \delta (z-b) \right] \right\rbrace f_{\sigma} = \]
\[
= \left( E - \frac{\hbar^{2} k^{2} }{2m}\right) f_{\sigma} \, .
\]
Here we have set $ V_{C} =0 $. The solution of this equation, which describes the bound states, is a piecewise smooth function 
\[
f(z) = \left\lbrace \begin{array}{ll}
A_{L} e^{\kappa_{L}(z-a)} & \textrm{at $ z< a $} ,\\
A_C e^{iqz} + B_C e^{-iqz} & \textrm{at $ a\leq z\leq b $}, \\
A_{R} e^{-\kappa_{R}(z-b)} & \textrm{at $ z> b $},
\end{array} \right. \, 
\]
whose derivative is discontinuous at the boundaries:
\[ \frac{d f}{d z}\vert_{z=a+0} - \frac{d f}{d z}\vert_{z=a-0} = - \sigma k \frac{V_{L}}{2mc^{2}} f(a) \, ,
\frac{d f}{d z}\arrowvert_{z=b+0} - \frac{d f}{d z}\arrowvert_{z=b-0} = \sigma k \frac{V_{R}}{2mc^{2}} f(b) \, . \]
The incoming parameters $ \kappa_{L,R} $ in the nonrelativistic approximation are defined in  Eq.  (\ref{kapa_LR}).

Matching conditions for the function at the boundaries  $ a $ and $ b $ lead to the condition (\ref{cond1m}), from which the bound states can be determined, and to Rashba dispersion law (\ref{calE_eps}) for 2D electron bands and spin vector   $ \mathbf{S} $, which are determined in Eq. (\ref{SiOM_eps}). In this case the upper spinor (large component) of the bispinor (\ref{varPsi_eps}) plays the role of the wavefunction.  

Therefore, Schr\"{o}dinger equation  (\ref{SEq}) gives only the solution, which corresponds to the given electric spin polarization, while, according to DE,  2D electrons, captured by the QW, can have different spin states,  which differ by the energy spectrum and by the electron spin orientation. At the given electric spin polarization the vector  $ \mathbf{S} $ lies in the plane of the potential layer and is 'bound' to the direction of the electron momentum, being perpendicular to it.  In the states with the given magnetic spin polarization  electron spins are oriented perpendicular to the layer, and, finally, in the states with the given spin polarization   $ \hat{\bm{\mathcal{S}}} $  they lie in the plane of the potential layer and are oriented along the chosen direction. The difference of the energy spectra of these three states, which is absent in the homogenous isotropic space, is, in fact, a sequence of the 'spin-orbit interaction'. Realization ('preparation') of one of these states is determined by external conditions, such as presence of electric and/or magnetic field, external pressure, interface properties, etc., and, therefore, should be manifested in various physical experiments.   

The reason for the origin  of SOI is determined by the fact that DE, unlike its nonrelativistic limit, does not admit separation of spatial and spin coordinates. This conclusion follows not only from the circumstance that even in the homogenous isotropic space spin operator   $ \mathbf{S} $ is not an integral of motion, but also from the fact that spin operators (\ref{S_p})-(\ref{genS}), which allow to classify the states by their spin degree of freedom, include also coordinate dependence since they contain the momentum operator.  

In conclusion, we stress that the joint procedure of the transition to the nonrelativistic limit is performed without the preliminary classification of the spin eigenstates and is based on the assumption that the down spinor is a small component of the bispinor for the states with positive energy \cite{Bete, LandauIV, Davydov}. As we have shown above, such assumption is valid for the states with the given electric spin polarization determined by the bispinor in Eq. (\ref{varPsi_eps}), only. In the general case the large and small components can belong to all four components of the bispinor, as it is in the case of spinors   (\ref{VarPsi_+})-(\ref{VarPsi_-}) and (\ref{VarPsi_+s})-(\ref{VarPsi_-s}). Probably, because of this fact the solution of Eq.  (\ref{SEq}) coincides with the solution of DE for the given electric spin polarization, but does not have the solutions corresponding to the given magnetic spin polarization  $ \bm{\mu }$, or spin polarization   $ \hat{\bm{\mathcal{S}}} $. Derivation of the nonrelativistic equations which take these facts into account, is a subject of a separate study and will be done elsewhere.  

\vspace{1cm}
{\textbf{Acknowledgement}} The work is done within the Fundamental Research Programme of the National Academy of Sciences of Ukraine.


\begin{thebibliography}{99}
\bibitem{Fabian}
J. Fabian, A. Matos-Abiague, C. Ertler, P. Stano, and I. \v{Z}uti\'{c}, Acta Phys. Slov. \textbf{57}, 565 (2007)).
\bibitem{Rashba1}
E. I. Rashba, Fiz. Tverd. Tela \textbf{2}, 1224 (1960) [Sov. Phys. Solid State \textbf{2}, 1109 (1960)]. 
\bibitem{Rashba2}
Y. A. Bychkov and E. I. Rashba, Pisma Zh. Eksp. Teor. Fiz. \textbf{39}, 66 (1984) [Sov. Phys. JETP Lett. \textbf{39}, 78 (1984)] .
\bibitem{Dirac}
P. A. M. Dirac, Proc. Roy. Soc., \textbf{A 117}, 610 (1928).
\bibitem{Bete}
Hans A. Bete. \emph{Intermediate Quantum Mechanics}, W. A. Benjamin, Inc. New York-Amsterdam, 1964.
\bibitem{LandauIV}
V.B. Berestetskiy, E.M. Lifshitz, L.P. Pitaevskiy,  \emph{Relativistic Quantum Theory}, part I (in Russian). Moscow, Nauka, 1968. 
\bibitem{Davydov}
O.S. Davydov.  \emph{Quantum Mechanics} (in Ukrainian), Vydavnychyi Dim Akademperiodyka, Kyiv, 2012. 
\bibitem{Klein}
O. Klein, Z. Phys. \textbf{53}, 157 (1929).
\bibitem{Dombey}
A. Calogeracos, N. Dombey, Int. J. Mod. Phys. A, \textbf{14}, 631 (1999).
\bibitem{Krekora}
P. Krekora, Q. Su, and R. Grobe, Phys. Rev. Lett. \textbf{92}, 040406 (2004); \textbf{93}, 043004 (2004). 
\bibitem{Gil}
P. Alberto, C. Fiolhais, and V. M. S. Gil, Eur. J. Phys. \textbf{17}, 19-24 (1996).
\bibitem{Leo}
S. De Leo and S. Giardino, J. Math. Phys. \textbf{55}, 022301 (2014).
\bibitem{Sokolov}
A.A. Sokolov, I.M. Ternov, \emph{Relativistic Electron} (in Russian), Moscow, Nauka, 1974
\bibitem{Bernardes}
E.S. Bernardes, J. Schliemann, M. Lee, J.C. Egues, and D. Loss, Phys. Rev. Lett. \textbf{99}, 076603 (2007). 

\end{thebibliography}
\end{document}